\documentclass[aip, jmp, amsmath,amssymb,preprint,reprint,author-year,author-numerical]{revtex4-1}

\usepackage[dvips]{graphicx}
\usepackage{dcolumn}
\usepackage{bm}

\begin{document}

\title{Hidden relationship between the electrical conductivity and the Mn 2$p$ core-level photoemission spectra 
in La$_{1-x}$Sr$_x$MnO$_3$}

\author{T. Hishida}
\email[] {t-hishida@mg.ngkntk.co.jp}
\author{K. Ohbayashi}
\affiliation{NGK SPARK PLUG CO., LTD., 2808 Iwasaki, Komaki, Aichi 485-8510, Japan}

\author{T. Saitoh}
\affiliation{Department of Applied Physics, Tokyo University of Science, Shinjuku, Tokyo 162-8601, Japan}

\date{\today}

\begin{abstract}
Core-level electronic structure of La$_{1-x}$Sr$_x$MnO$_3$ has been studied by x-ray photoemission spectroscopy (XPS). We first report, by the conventional XPS, the well-screened shoulder structure in Mn 2$p_{3/2}$ peak, which had been observed only by hard x-ray photoemission spectroscopy so far. Multiple-peak analysis revealed that the Mn$^{4+}$ spectral weight was not proportional to the nominal hole concentration $x$, indicating that a simple Mn$^{3+}$/Mn$^{4+}$ intensity ratio analysis may result in a wrong quantitative elemental analysis. Considerable weight of the shoulder at $x$=0.0 and the fact that the shoulder weight was even slightly going down from $x$=0.2 to 0.4 were not compatible with the idea that this weight simply represents the metallic behavior. Further analysis found that the whole Mn 2$p_{3/2}$ peak can be decomposed into four portions, the Mn$^{4+}$, the (nominal) Mn$^{3+}$, the shoulder, and the other spectral weight located almost at the Mn$^{3+}$ location. We concluded that this weight represents the well-screened final state at Mn$^{4+}$ sites, whereas the shoulder is known as that of the Mn$^{3+}$ states. We found that the sum of these two spectral weight has an empirical relationship to the conductivity evolution with $x$.
\end{abstract}



\maketitle

\section{INTRODUCTION}

	3$d$ transition-metal (TM) oxides with the perovskite-type structure $R_{1-x}A_xM$O$_3$ ($R$: rare earths, $A$: 
alkaline earths, $M$: TM)  exhibit a variety of physical properties. This is because complex interplays among
the charge, spin and orbital degrees of freedom exist in these compounds, as often discussed. \cite{Mandal00,Ohtani00,Hemberger02}
However, what are more important than the interplays themselves are subtle balances of them.
In this sense, $R_{1-x}A_xM$O$_3$ oxides have a big advantage because the tertiary formula of their compounds 
($RM$O$_3$ and $AM$O$_3$) enables us to finely tune the balances, primarily the number of charge carriers, 
by controlling $x$ with minimal structural/environmental changes at the TM sites.
	In many 3$d$ TM perovskite-type oxides, a family of manganese oxide $R_{1-x}A_x$MnO$_3$ and their relatives occupies 
a special place because of its extraordinary property, the colossal magnetoresistance (CMR).\cite{vonHelmolt93,Tokura94,Ramirez97}
This phenomenon is widely known as a manifestation of the complex interplays that are emphasizing a strong coupling
between metallic conductivity and ferromagnetism, although it has not reached complete understanding yet.
CMR is thus still attracting much attention and interest by a number of researchers. 

To seek the origin of CMR, particularly an aspect of the electrical conductivity, x-ray photoemission spectroscopy (XPS) is one of the powerful tools because it directly probes the electrons coming out of the target sample.\cite{Chainani93,Mannella06,Matsuno02}  Core-level XPS is known as a typical element-selective technique and suitable for investigating the valence state of TM ions in the sample. However, the photoemission spectroscopy is generally surface sensitive and one should always pay appropriate attention when comparing the obtained spectra with macroscopic properties such as bulk electrical conductivity. Bulk-sensitive hard x-ray photoemission spectroscopy (HX-PES)\cite{Kobayashi03}  using synchrotron radiation was a breakthrough to overcome this difficulty and an increasing number of HX-PES studies have been reported in the past few years.\cite{Horiba04,Tanaka06,Offi08,Ueda09}
	Among them, Horiba {\it et al.}\cite{Horiba04} found a new shoulder structure in the lower binding energy side of Mn 2$p_{3/2}$ core-level HX-PES peak of La$_{1-x}$Sr$_x$MnO$_3$ (LSMO) thin films, which had never been observed until then because, according to their report, surface sensitivity in the conventional photoemission measurements is higher than HX-PES. More interestingly, this new structure changed its intensity with metal-insulator transition (MIT) either by temperature or hole concentration $x$, being interpreted as a well-screened peak due to electronic states at/near Fermi energy ($E_{\rm F}$).
	With regard to the temperature-dependent change, Ueda \textit{et al.} found a good agreement between the conductivity and the magnetization calculated using a hybridization term deduced from the core-level spectra of their La$_{0.85}$Ba$_{0.15}$MnO$_3$ thin films.\cite{Ueda09}
	However, the origin of the new structure and its relation to electrical conductivity has not yet been clarified completely because (1) the insulating LaMnO$_3$ (LMO) also has this structure as a hump and (2) no systematic investigations extending for the Sr-rich region have been done so far.

In this paper, we re-examine the Mn 2$p$ core-level photoemission spectra of sintered bulk LSMO by conventional XPS to search how spectral features, including the above-mentioned shoulder structure, evolve with $x$ and how they can be correlated to electrical conductivity. From a detailed multiple-peak analysis, we propose an empirical relationship between some components of Mn 2$p_{3/2}$ peak and the electrical conductivity.

\section{Experimental}

Polycrystalline sample of La$_{1-x}$Sr$_x$MnO$_3$ ($x$=0.0, 0.1, 0.2, 0.33, 0.4, 0.5, 0.55, 0.67, 0.8, 0.9, and 1.0) were prepared by solid-state reaction. Starting powders of La(OH)$_3$ (99.9 \%, Shin-Etsu Chemical Co., Ltd.), SrCO$_3$ (99.8 \%, SAKAI CHEMICAL INDUSTRY Co., Ltd.), and Mn$_2$O$_3$ (99.9 \%, KOJUNDO CHEMICAL LABORATORY Co., Ltd.) were weighted in specific proportions and then mixed in a ball milling for 15 hr using zirconia balls and ethanol. The mixed powders were dried and heated at 1100 $^{\circ}$C for 2 hr in air. After it had cooled, they were crushed in the ball milling again for 15 hr. The powders were dried, and pressed into pellet forms under the isostatic pressure of 0.8 ton/cm$^2$, and then sintered at 1600 $^{\circ}$C for 1hr in air. The crystal structure of LSMO was characterized by Rietveld treatment of the x-ray diffraction (XRD) profiles in Table~\ref{table1}. XRD data were measured at the BL19B2 of SPring-8.

	The XPS measurements were performed with a PHI Quantera SXM (base pressure 5$\times$10$^{-9}$ Torr) instrument using a monochromatic Al K$_{\alpha}$ source ($h\nu$=1486.6 eV) with the total energy resolution of about 0.64 eV in full width at half maximum (FWHM). The analyzer pass energy was set to 55 eV for narrow scans. In order to obtain fresh, clean surfaces, the samples were fractured in the prep chamber at room temperature right before the measurements. The binding energy was corrected by using the value of 84.0 eV for the Au 4$f_{7/2}$ core-level spectrum. The measurement vacuum was better than 1$\times$10$^{-8}$ Torr.
 Electrical conductivity was measured by a dc four probes method (ADVANTEST, Type 6242).\cite{Pauw58}

\section{Results and discussions}

\begin{table}[b]
\caption{\label{table1} Crystal structure, space group, and lattice parameters of LSMO. }
\begin{ruledtabular}
\begin{tabular}{cllccc}
		&\multicolumn{1}{c}{Crystal}		&\multicolumn{1}{c}{Space}	& \multicolumn{3}{c}{Lattice parameters (\AA) }	\\
$x$	&\multicolumn{1}{c}{structure}	&\multicolumn{1}{c}{group}	&$a$		&$b$		&$c$	\\
\tableline
0.00	&orthorhombic		&$Pnma$			&5.63445	&7.73557	&\phantom{0}5.53516	\\
0.10	&orthorhombic		&$Pnma$			&5.76259	&7.75083	&\phantom{0}5.54745	\\
0.33	&rhombohedral		&$R$\={3}$c$		&5.50074	& 				&13.35744					\\
0.40	&rhombohedral		&$R$\={3}$c$		&5.48504	& 				&13.34683					\\
0.50	&tetragonal			&$I4/mcm$			&5.44390	&				&\phantom{0}7.75818	\\
0.67	&tetragonal			&$I4/mcm$			&5.42619 	&				&\phantom{0}7.70060	\\
0.80	&cubic					&$Pm$\={3}$m$	&3.82500	&				&								\\
1.00	&hexagonal				&$P6_3/mmc$		&5.44578	&				&\phantom{0}9.07628	\\
\end{tabular}
\end{ruledtabular}
\end{table}

\begin{figure}[b]
	\begin{center}
 	\includegraphics[width=80mm,keepaspectratio]{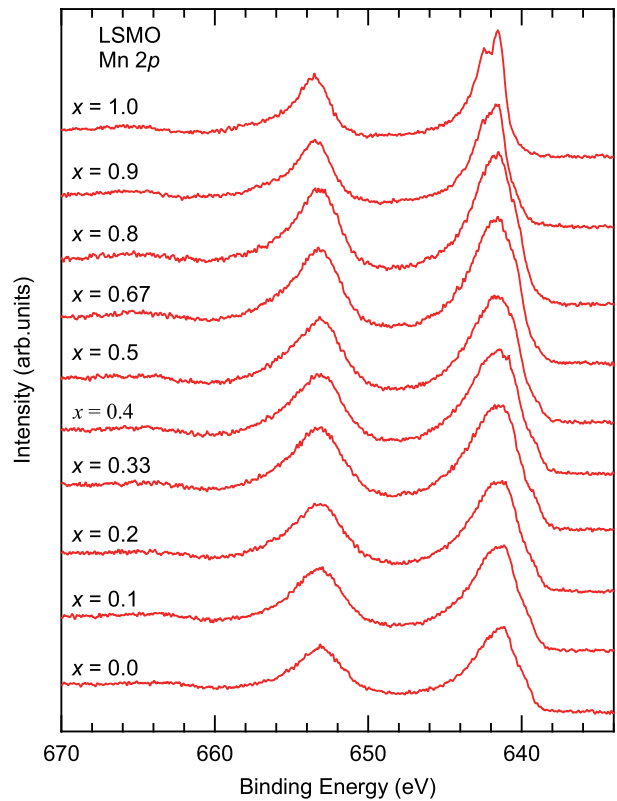}
	\caption{Mn 2$p$ core-level XPS spectra of LSMO with various Sr concentrations.}
	\label{Mn2pwhole}
	\end{center}
\end{figure}

Mn 2$p$ core-level XPS spectra of LSMO with various Sr concentrations are shown in Fig.~\ref{Mn2pwhole}. All the spectra have 2$p_{3/2}$ and 2$p_{1/2}$ spin-orbit doublet peaks located at $\sim$641 eV and $\sim$653 eV, respectively. The broad peaks located around 665$-$666 eV are the charge-transfer satellites of the 2$p_{1/2}$ peak.\cite{Saitoh95,Zampieri02}
	The lineshape of Mn 2$p_{3/2}$ of LMO is essentially the same as that of Mn$_2$O$_3$ with the same 3$d^4$ (Mn$^{3+}$) electron configuration (see Fig.~\ref{Mn2O3MnO2} (a)), except that the LMO peak has a small shoulder, which is obvious up to $x$=0.5. We will discuss this shoulder in detail later.
	On the other hand, the SrMnO$_3$ (SMO) spectrum, having a double-peak structure characteristic of 3$d^3$ electron configuration, is almost identical to that of MnO$_2$ with the 3$d^3$ (Mn$^{4+}$) electron configuration (see Fig.~\ref{Mn2O3MnO2} (b)).
	The Mn 2$p_{3/2}$ peak width gradually increases probably due to the chemical shift of the Mn 2$p_{3/2}$ peak from Mn$^{3+}$ to Mn$^{4+}$, indicating that the Mn$^{4+}$ component systematically grows with $x$ as expected. However, it actually increases even up to $x$=0.8 and then rapidly decreases in $x>0.8$ with significant changes in lineshape. Hence, the Mn valence may be changing more rapidly in $x>0.8$ than $x \leq 0.8$.

\begin{figure}[t]
	\begin{center}
 	\includegraphics[width=80mm,keepaspectratio]{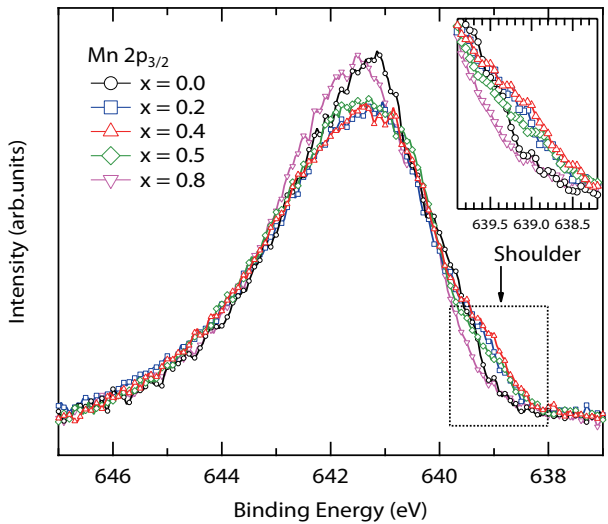}
	\caption{(Color online) Mn 2$p_{3/2}$ core-level XPS spectra of LSMO with various Sr concentrations.}
	\label{Mn2p3-2}
	\end{center}
\end{figure}

	Figure~\ref{Mn2p3-2} shows detailed comparisons in the Mn 2$p_{3/2}$ region. Here, the background intensity due to secondary electrons were subtracted from the spectra using the Shirley-type background correction\cite{Shirley72} and the spectra have been normalized with their areas from 635.0 to 648.5 eV.
	One can observe that the shoulder structure located at $\sim$639 eV already exists in $x$=0.0,  develops upto $x=0.4$ and then loses its weight for $x>0.5$.
	This shoulder, characteristic of the LSMO system (particularly in the metallic phase), has been considered to be measurable only by HX-PES.\cite{Horiba04} We note that the present result is the first decisive observation of the shoulder in the spectra from sintered bulk samples by the conventional lab XPS, which is less bulk-sensitive than HX-PES. The intensity of the shoulder is, however, smaller than observed in HX-PES, probably due to less bulk sensitivity as already demonstrated.\cite{Horiba04}
	It is also noted that the shoulder weight is substantial even at the insulating $x$=0.0 and still finite upto $x$=0.8 (discussed later), although insulating Mn$_2$O$_3$ has no this weight (see Fig.~\ref{Mn2O3MnO2} (a)). 
	Interestingly, the shoulder was even observed in spectra from the surface fractured in air (not shown). Combining the fact that the O 1$s$ spectrum from the fractured surface in air has an extra surface-related peak, this may be indicating that the electronic state related to this shoulder has purely bulk nature and/or this electronic state at surface is very stable against oxidization.

	Since the finite shoulder weight at $x$=0.0 may be caused by mobile carrier (or Mn$^{4+}$) due to off stoichiometry,\cite{Horiba04} we have checked stoichiometry of our LMO sample. Several structural studies\cite{Urushibara95, vanRoosmalen95, Ritter97} reported that the stoichiometric LMO is orthorhombic and should have the cell volume of 243$-$244 \AA$^3$, which decreases with increasing $\delta$ of LaMnO$_{3+\delta}$ or $x$ of La$_{1-x}$Sr$_x$MnO$_3$ (up to $x<0.175$ (Ref.~\onlinecite{Urushibara95})). The cell volume of our LMO was 241.1 \AA$^3$, which is more stoichiometric than LaMnO$_{3.025}$ (5 \% Mn$^{4+}$) with its cell volume of 240.7 \AA$^3$ reported by Ritter {\it et al}.\cite{Ritter97}  Hence, our LMO sample was practically stoichiometric although it may have a small amount of Mn$^{4+}$, less than about 4 \% ($\delta$=0.02). In fact, theoretical calculations on eight MnO$_6$ clusters reproduced this shoulder in LMO, too.\cite{vanVeenendaal06}
	All these facts unambiguously demonstrate that the shoulder is not merely a consequence of doping-induced electronic states at $E_{\rm F}$ as discussed by Horiba {\it et al.}, who attributed the shoulder in a LMO thin film to excess oxygen.\cite{Horiba04}

\begin{figure}[t]
	\begin{center}
 	\includegraphics[width=70mm,keepaspectratio]{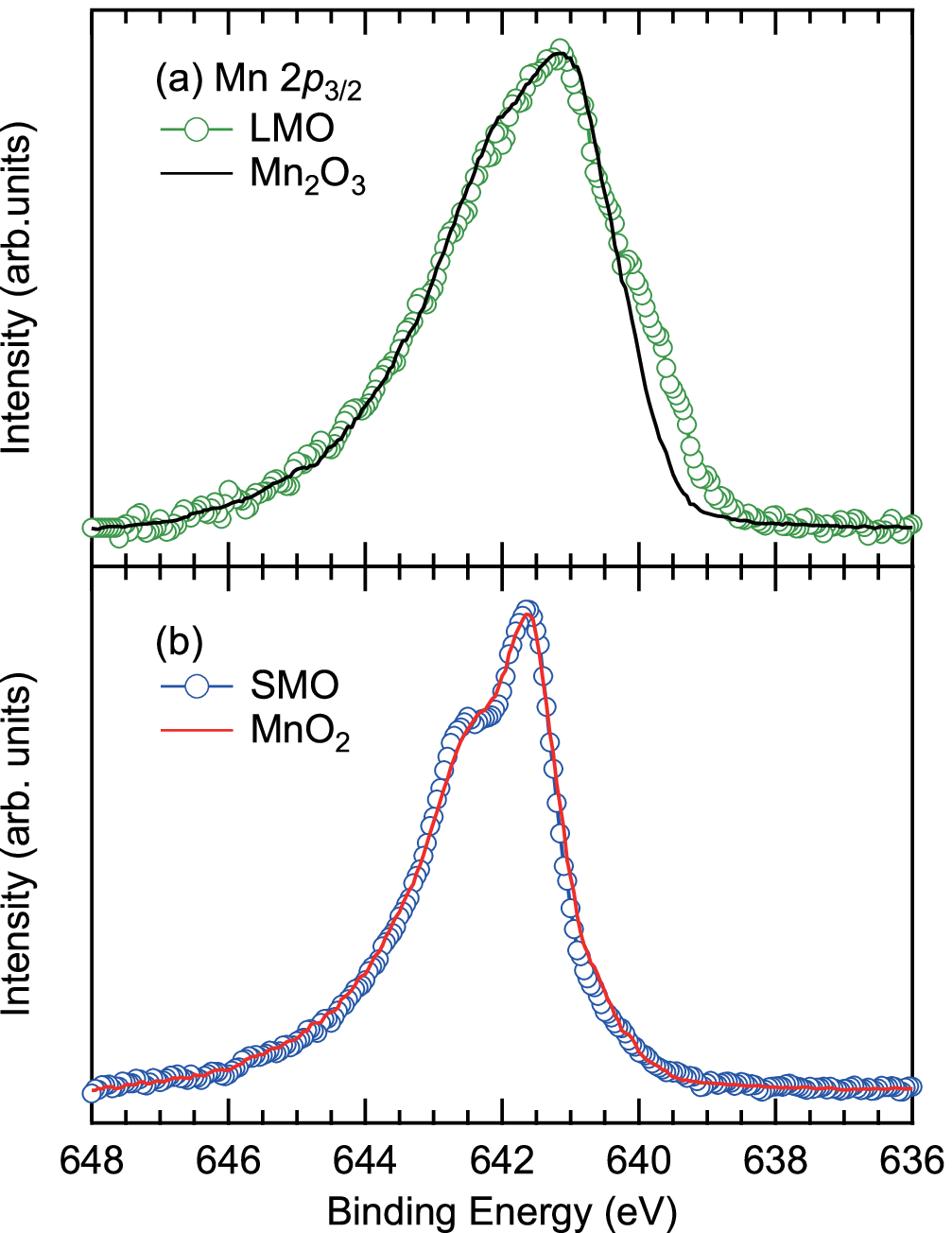}
	\caption{(Color online) Mn 2$p_{3/2}$ spectra of (a) LMO compared with Mn$_2$O$_3$ (Ref.~\onlinecite{Stranick99-1}) and (b) SMO compared with MnO$_2$.\cite{Stranick99-2}}
	\label{Mn2O3MnO2}
	\end{center}
\end{figure}
\begin{figure}[t]
	\begin{center}
 	\includegraphics[width=70mm,keepaspectratio]{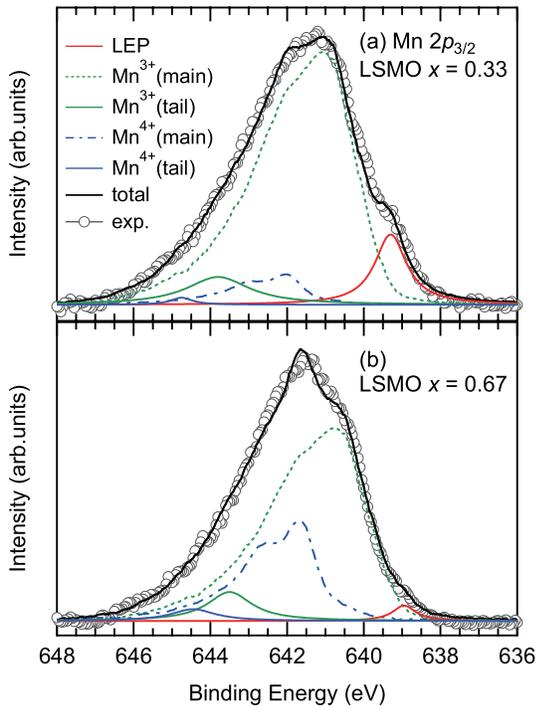}
  \caption{(Color online) Multiple-peak fitting result for Mn 2$p_{3/2}$ spectrum of (a) La$_{0.67}$Sr$_{0.33}$MnO$_3$ and (b) La$_{0.33}$Sr$_{0.67}$MnO$_3$.}
\label{Fitting}
\end{center}
\end{figure}

	In order to investigate the origin of the shoulder structure and extract more detailed information on the Mn valence evolution, we performed multiple-peak fitting for the Mn 2$p_{3/2}$ spectra in a different way from the literature, which has not been well-established so far but is accurate enough we believe to track the shoulder structure.
	In many reports, it is usually assumed that a spectrum of a certain $x$ can be reproduced by an appropriate superposition the Mn$^{3+}$ and Mn$^{4+}$ spectra.\cite{Kowlik10,Bindu11,Ogawa06,Nesbitt98} However, such a fitting won't be adequate here because the peak width and the shoulder intensity in particular do not evolve linearly with $x$.
	Hence, we adopt a Mn$_2$O$_3$ (Ref.~\onlinecite{Stranick99-1}) and MnO$_2$ (Ref.~\onlinecite{Stranick99-2}) spectrum as references of the both ends of the Mn$^{3+}$ and Mn$^{4+}$ states, respectively;
	Figure~\ref{Mn2O3MnO2} (a) shows that the Mn$_2$O$_3$ spectrum (shifted by no more than 0.1 eV towards $E_{\rm F}$) can explain the LMO spectrum well only except the shoulder.\cite{note2} On the other hand, with a very tiny shift of 0.05eV towards $E_{\rm F}$, the MnO$_2$ spectrum almost completely agrees with the SMO spectrum, as shown in Fig.~\ref{Mn2O3MnO2} (b).\cite{note2}
	The Mn$_2$O$_3$ and MnO$_2$ spectra are thus appropriate as the both ends in order to  reproduce the spectra in all range of $x$ and track the intensity change at the shoulder.
	For the intermediate $x$'s, we further introduced an additional peak for the Mn$^{3+}$ and Mn$^{4+}$ main peaks, respectively, to simulate the spectral weight in the higher binding energy tail of the Mn 2$p_{3/2}$ spectra. All the three peaks (including the one for the shoulder named ``LEP", which stands for this low-binding energy peak) were assumed to have the Voigt profile for simplicity.
	The binding energies and the intensities of the end spectra (Mn$_2$O$_3$ and MnO$_2$ ones with tiny shifts) were varied. Thus, the fitting parameters were these and the magnitudes, widths, and the binding energies of the three functions.
	To reduce ambiguity of the fitting, we subtracted the background intensity due to secondary electrons from all the spectra using the Shirley-type background correction.\cite{Shirley72}

\begin{figure}[t]
	\begin{center}
 	\includegraphics[width=77mm,keepaspectratio]{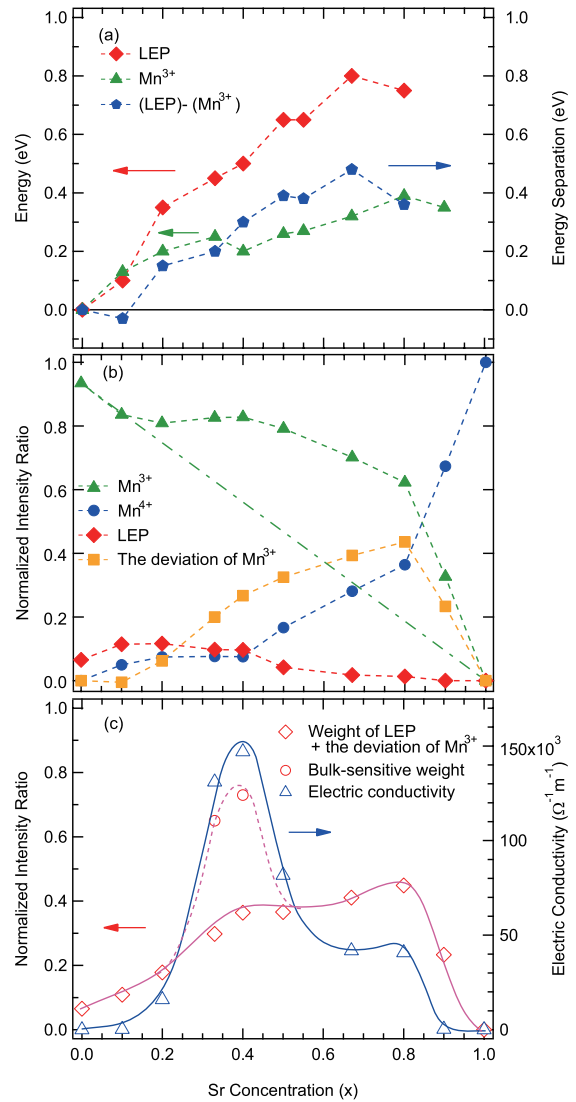}
	\caption{(Color online) (a) Binding energy shift of LEP and Mn$^{3+}$, and their separation plotted as a function of Sr concentration $x$. 
(b) Relative spectral weight of the decomposed peaks deduced from the multiple-peak analysis. Green triangles, blue circles, and red diamonds denote Mn$^{3+}$, Mn$^{4+}$, and LEP, respectively. The green dashed line denotes the Mn$^{3+}$ weight expected by $x$, namely the La$^{3+}$/Sr$^{2+}$ ratio (not including the LEP weight at $x$=0.0). Orange squares are deviation of the Mn$^{3+}$ weight from this expected Mn$^{3+}$/Mn$^{4+}$ ratio. 
(c) Electrical conductivity of LSMO at room temperature and the sum of weight of LEP and the deviation of Mn$^{3+}$. The open circles and dotted line show an expected behavior for an ideally bulk-sensitive measurement (see text).}
	\label{Fig5}
	\end{center}
\end{figure}

	Figure~\ref{Fitting} shows decomposition of the (a) $x$=0.33 and (b) $x$=0.67 spectra as examples of the fitting. It is now obvious that the low-binding energy hump in the experimental spectrum is not just a continuation of the main peak but does originate in LEP. In this way, we succeeded in tracking the weight of LEP upto $x$=0.8, beyond any literature, where it is too small to be noticed as a structure in the spectrum before decomposition. In fact, our experimental spectrum of $x$=0.8 is in good agreement with a recent report by Bindu {\it et al.}, although their fitting did not include our LEP as an independent peak.\cite{Bindu11}

	The binding energy shifts of the Mn 2$p_{3/2}$ components with $x$ are shown in Fig.~\ref{Fig5}~(a). The Mn$^{3+}$ main peak slightly moves towards $E_{\rm F}$ with $x$ in agreement with literature\cite{Horiba04}, while LEP moves in the same direction more rapidly until $x$=0.67 and then back away from $E_{\rm F}$. As a consequence, the separation between the Mn$^{3+}$ main peak and LEP increases with $x$ until $x$=0.67, but eventually decrease. This trend agrees with what was observed in the thin films by Horiba {\it et al.} although their data were limited up to $x$=0.55.\cite{Horiba04} In their MnO$_6$ cluster-model analysis, the separation was scaled by the hybridization term $V^*$ between the Mn 3$d$ states and the doping-induced $E_{\rm F}$ state, leading them to the conclusion that $V^*$ represents the magnitude of screening by conduction electrons (We will denote this screening as``conduction electron screening" (CES)). However, this cannot explain the fact that there is substantial spectral weight at the insulating LMO unless an amount of conduction electrons are assumed. Although they attributed it to excess oxygen, it is somewhat questionable whether such large peak in both their thin-film and our bulk samples can satisfactorily be explained by this scenario. On the other hand, van Veenendaal's calculation using a Mn$_8$ cluster also reproduced a new peak corresponding to our LEP, but he predicted the opposite behavior of the separation with $x$.\cite{vanVeenendaal06} In his framework, this peak represents nonlocal screening (NLS) effects and does not necessarily correspond to the well-screened final state due to existence of coherent band at $E_{\rm F}$.

	Figure~\ref{Fig5}~(b) summarize the relative spectral weight of the decomposed peaks deduced from the multiple-peak analysis. 
	It is found that the Mn$^{3+}$ spectral weight does not linearly decrease with $x$ but is always higher than the expected Mn$^{3+}$/Mn$^{4+}$ ratio; it is almost constant in $0.1 \leq x \leq 0.4$, gradually decreases with further increase in $x$ until $x$=0.8, and finally rapidly decreases to zero in $x>0.8$. On the other hand, the Mn$^{4+}$ spectral weight is almost constant in $0.1 \leq x \leq 0.4$ and rapidly increases after that in $x \geq 0.4$. Even if we ignore weight of LEP (it is not very large; see below), the above result is clearly demonstrating that the simple Mn$^{3+}$/Mn$^{4+}$ ratio assumption based on the La$^{3+}$/Sr$^{2+}$ ratio is not appropriate. After discussing LEP, we will be back to this issue later.

	The relative spectral weight of LEP first increases from $x$=0.0 to 0.2, slightly decreases until $x$=0.4, and then goes down more rapidly towards $x$=1.0. This change is, again, essentially the same as what was observed in thin films both at 300 K and 40 K as represented by the evolution of $V^*$, having the maximum at $x$=0.2.\cite{Horiba04} Hence, the CES scenario is not very compatible with the electrical conductivity change with $x$, which is found in vast literature.\cite{Urushibara95} We also show our data at 300 K in Fig.~\ref{Fig5}~(c). Nevertheless, their temperature-dependent intensity changes in the peak at each $x$ does demonstrate the importance of CES.
	A simple NLS scenario neither explains the spectral weight change with $x$ well since the calculated NLS peak behavior, moving towards the main peak and rapidly losing its weight, is not what was observed in the experiments.
	However, his calculations also predicted that the A-type antiferromagnetism (A-AF) coupled to the Jahn-Teller (JT) distortion at $x$=0.0 substantially reduces the separation and the spectral weight.\cite{vanVeenendaal06}

	A promising  interpretation of our and past experiments/theories is that LEP includes both NLS and CES nature of Mn$^{3+}$ sites; at LMO ($x$=0.0), the A-AF coupled to the JT distortion first suppresses the NLS peak.\cite{note1} The remnant of the A-AF and the JT distortion upto $x$$\sim$0.1 keeps suppressing the NLS peak to some extent, but this suppression decreases with decreasing A-AF and JT effects.
	In other words, the NLS shoulder affected by the A-AF and JT effects at $x$=0.0 continuously evolves to a larger NLS peak in the ferromagnetic metallic phase at $x$=0.1$\sim$0.2, and then loses its weight simply due to decrease in $e_g$ electrons in $x\geq 0.2$. In this metallic range, however, CES nature develops in parallel, preventing from a rapid weight loss of the peak. Finally, both NLS and CES weight disappear with decrease in $e_g$ electrons.
	The change of the separation from the main peak with $x$ can also be understood by the same idea; with decreasing A-AF and JT effects with $x$, the separation becomes large.\cite{vanVeenendaal06} In the range of $x\geq 0.2$, increasing CES nature takes over the origin of the increase in the separation.\cite{Horiba04} Finally, the separation turns to decrease a little in the largest $x$'s because of dominant NLS nature.\cite{vanVeenendaal06}
	Here we note that the NLS and CES cannot be distinguished in the metallic phase.

	Based on the above interpretation, we may expect that the spectral weight of LEP behaves as the electrical conductivity with $x$. Fig.~\ref{Fig5}~(c) shows Sr-doping dependence of electrical conductivity at 300 K. It increases with $x$ in the range of $0.0 \leq x \leq 0.4$, taking the maximum at $x=0.4$, gradually decreases in $0.4 \leq x \leq 0.8$, and a rapid drop follows in $x \leq 0.8$.
	However, this behavior does not agree with that of LEP described earlier. In particular, the conductivity of LMO ($x$=0.0) is lower than $x$=0.8 while the LEP weight at $x$=0.0 is much larger than at $x$=0.8. Hence, the spectral weight of LEP alone is not actually representing electrical conductivity.

\begin{figure}[t]
	\begin{center}
 	\includegraphics[width=80mm,keepaspectratio]{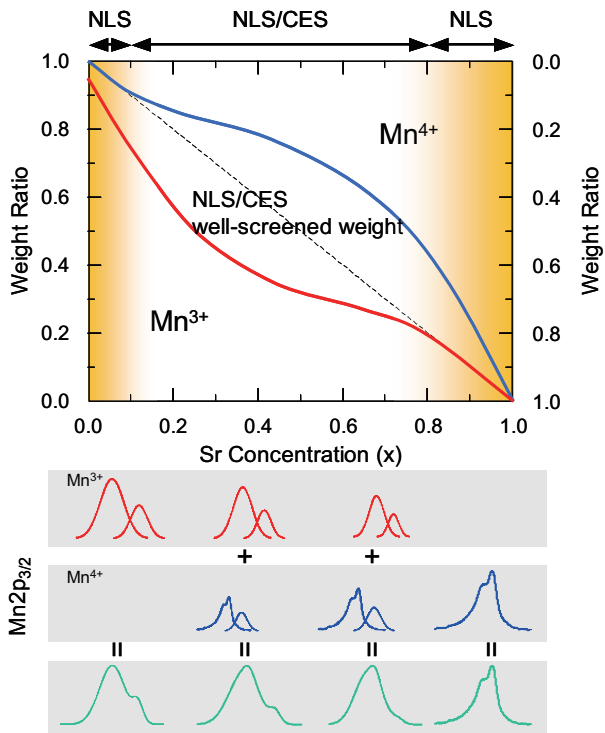}
  \caption{(Color online) Schematic weight decomposition of Mn 2$p_{3/2}$ peak. The area below (above) the red (blue) curve denotes the Mn$^{3+}$ (Mn$^{4+}$) main peak. The well-screened peak intensity due to NLS and/or CES is in between. Note that this figure does not include the temperature-dependent changes.}
\label{Mn2pSchematic}
\end{center}
\end{figure}

	Why does not the intensity variation of the well-screened peak due to NLS/CES represent the electrical conductivity?
We believe that this question is closely related to the deviation of the Mn$^{3+}$ weight from the expected value; both NLS\cite{vanVeenendaal06} and CES\cite{Horiba04} calculations were limited to Mn$^{3+}$ sites, ignoring these effects at Mn$^{4+}$ sites. However, there should be such well-screened peak for the Mn$^{4+}$ configuration, too and it should be located at a lower binding energy side of the Mn$^{4+}$ main peak. Hence the Mn$^{4+}$ NLS/CES peak would be overlapping with the Mn$^{3+}$ main peak and this is the reason why the apparent Mn$^{3+}$ weight is larger than expected, namely the observed deviation of the Mn$^{3+}$ weight probably represents the Mn$^{4+}$ NLS/CES weight.

	The sum of weight of LEP and the deviation (orange squares in Fig.~\ref{Fig5}~(b)) is compared with the electrical conductivity in Fig.~\ref{Fig5}~(c).
The overall trend of the weight sum has fair agreement with that of the electrical conductivity.
	From $x$=0.0 to 0.2, the weight sum increases with $x$ in agreement with the conductivity.\cite{note3}   It even reproduces the smaller conductivity of $0.0 \leq x \leq 0.2$ than $0.67 \leq x \leq 0.8$ in spite of the larger LEP weight in the former range. The conductivity has a peak at $x$=0.4 and a large drop from $x$=0.4 to 0.5 in agreement with other reports,\cite{Hemberger02} which may be reflecting of a structural phase transition here.\cite{Hemberger02,Chmaissem03}
	However, the weight sum fails to reproduce the peak and the drop from $x$=0.4 to 0.5 in the conductivity.
This is probably because the lab-XPS measurement is more surface sensitive than HX-PES and the bulk-sensitive CES weight\cite{Horiba04} is underestimated in the metallic range of $x$ due to reduced ferromagnetism at surface.\cite{JHPark98}
	In other words, if the spectra were ideally bulk-sensitive, the weight sum of the $x$=0.33 and 0.4 (which are the most metallic samples) would be larger than observed and the weight sum behavior will be more similar to the electric conductivity, namely, the weight sum qualitatively reproduces the peak and the drop and even the plateau of the conductivity around  $0.67 \leq x \leq 0.8$ as well. 

	A schematic relative spectral weight of each component in Mn $2p_{3/2}$ core-level peak deduced from the present results is shown in Fig.~\ref{Mn2pSchematic}. The total NLS/CES weight comes from both Mn$^{3+}$ and Mn$^{4+}$ sites and
the NLS nature is dominant in the well-screened spectral weight near the both insulating ends, while the CES nature becomes dominant in the metallic phase. However, we note that even NLS will not be observed if the system is very insulating, implying that NLS in the insulating phase still manifests the electrical conductivity to some extent. As a result, the total NLS/CES weight is found to be a rough measure of electrical conductivity of the system.


\section{Conclusions}

	In conclusion, we have studied the core-level electronic structure of La$_{1-x}$Sr${_x}$MnO$_3$ by XPS. Employing a careful sample synthesis, we have, for the first time by the conventional XPS, succeeded in observing a clear well-screened structure at the lower binding energy side of the Mn 2$p_{3/2}$ core-level spectra, which had been observed only by HX-PES so far. By the multiple-peak analysis on the Mn 2$p_{3/2}$, we identified the well-screened peak in a wide range of $0.0 \leq x \leq 0.8$ and established its intensity/location changes with $x$, which were not consistent with a simple metallic screening. The multiple-peak analysis also revealed that the Mn$^{4+}$ spectral weight was not proportional to the nominal hole concentration $x$ but substantially less than expected. The resultant excess Mn$^{3+}$ was considered as the well-screened peak from the Mn$^{4+}$ sites, which was supported by an empirical relationship between the sum of the well-screened weight and the electrical conductivity.

\section*{ACKNOWLEDGEMENTS}
We would like to thank H. Kozuka for providing us with LSMO samples used in this work. We are also grateful to Dr. T. Ida for determining the lattice parameters. The synchrotron radiation experiments were performed at SPring-8 with the approval of the Japan Synchrotron Radiation Research Institute (JASRI) (Proposal No. 2011B1621 and 2011B1622).


\begin{thebibliography}{99}
\bibitem{note2} In order to compare the line shape and highlight the shoulder structure, the intensity of LMO and Mn$_2$O$_3$ was normalized with its maximum peak height. So is the case of SMO and MnO$_2$.
\bibitem{note1}No NLS peak in Mn$_2$O$_3$ could be explained by its insulating nature while the ferromagnetic planes in A-AF of LMO is on the verge of metal due to the double exchange mechanism. In this sense, the NLS peak may have a small amount of CES nature.
\bibitem{note3}The agreement will be further improved if we subtract the LEP weight of LMO as an insulating NLS peak.
\end{thebibliography}
\end{document}